\documentclass[aps,prb,twocolumn,longbibliography]{revtex4-2}
\usepackage{amsmath,amssymb,graphicx}
\usepackage{tikz}
\usepackage{dsfont}
\usepackage[utf8]{inputenc}
\newcommand\bea{\begin{eqnarray}}
\newcommand\eea{\end{eqnarray}}
\newcommand\beq{\begin{equation}}
\newcommand\eeq{\end{equation}}

\def\al{\alpha}

\begin{document}

\title{Josephson effect in bipolar magnetic semiconductors}
\author{Polireddi Naveen}
\affiliation{School of Physics, University of Hyderabad, Prof. C. R. Rao Road, Gachibowli, Hyderabad 500046, India}

\author{Abhiram Soori}
\email{abhirams@uohyd.ac.in}
\affiliation{School of Physics, University of Hyderabad, Prof. C. R. Rao Road, Gachibowli, Hyderabad 500046, India}

\begin{abstract}
We theoretically investigate equilibrium currents in a one-dimensional Josephson junction consisting of a bipolar magnetic semiconductor (BMS) in the middle. We show that the intrinsic exchange splitting of the spin-resolved bands enables purely electrical control of the $0$--$\pi$ transition through gate-tunable modulation of the BMS chemical potential, eliminating the need for an external magnetic field. This provides a viable route toward electrically tunable $\pi$-junction behavior and highlights the potential of BMS-based Josephson devices for phase-controllable superconducting electronics. Furthermore, in the presence of Rashba spin-orbit coupling, we find an anomalous Josephson effect characterized by a finite equilibrium supercurrent at zero phase difference. This behavior originates from the intrinsic breaking of inversion and time-reversal symmetry associated with the spin-polarized electronic structure of the BMS with spin-orbit coupling. Interestingly, despite the simultaneous breaking of time-reversal and inversion symmetries---conditions often associated with nonreciprocal superconducting transport--we do not observe a Josephson diode effect. Our results therefore highlight an important distinction between anomalous Josephson transport and superconducting nonreciprocity: the former does not necessarily imply a finite critical-current asymmetry between opposite current directions. 
\end{abstract}

\maketitle

\section{Introduction}

Placing a metallic or thin insulating layer between two superconductors with distinct phases results in the flow of a dissipationless equilibrium current, a phenomenon identified as the Josephson effect~\cite{Josephson}. The current--phase relation (CPR) captures how this supercurrent depends on the phase gradient across the junction~\cite{CPR}. Furthermore, in specific Josephson junctions where time-reversal and inversion symmetries are broken, a finite equilibrium current can exist even when the phase difference vanishes, which is termed the anomalous Josephson effect~\cite{an.joseph}.

Bipolar magnetic semiconductors (BMSs) have recently garnered attention as a unique family of magnetic materials characterized by a finite band gap separating valence and conduction bands with opposite spin polarizations. This unique spin-resolved band topology naturally induces an intrinsic exchange splitting. As a result, time-reversal symmetry is inherently broken in BMSs, making them an ideal platform for exploring spintronic effects in the absence of external magnetic fields~\cite{BMS1,BMS2,BMS3,BMS4,BMS5,BMS6,BMS7,naveen2026car,Suresh2021}.

Superconductor/ferromagnet (S/F) and S/F/S junctions have been thoroughly investigated, largely because of their capacity to host $0$--$\pi$ transitions~\cite{ferromagnet1,ferromagnet2,ferromagnet3,ferromagnet4,ferromagnet6,ferromagnet7}. In junctions harboring two ferromagnetic layers, altering their relative magnetization orientation serves as a mechanism to trigger this $0$--$\pi$ crossover~\cite{ferromagnetspinvalve}. Interfacing a standard superconductor with a ferromagnet induces a proximity effect wherein the electrons of opposite spins have different momenta. This momentum disparity causes an oscillatory spatial dependence of the Cooper-pair correlation function inside the ferromagnetic region~\cite{ferromagnet5}. The system minimizes its energy at a phase difference of $0$ in the $0$-state, and at $\pi$ in the $\pi$-state. Moreover, $\pi$-junctions are heavily utilized in the architecture of superconducting qubits and quantum circuits, leveraging the built-in $\pi$ phase shift for enhanced control and tuning of circuit functionalities.

Josephson junctions utilizing van der Waals heterostructures offer the distinct benefit of atomically precise, ultrathin interfaces, which are highly suited for thickness-controlled $0$--$\pi$ transitions~\cite{vanderwaal}. 
In parallel, the anomalous Josephson effect has been successfully realized in proximity-induced superconducting junctions by employing semiconducting nanowires with Rashba spin--orbit coupling (SOC) under an appropriately aligned magnetic field~\cite{an.joseph.semi}.
A similar anomalous response has also been investigated in ferromagnet--SOC junctions, where the synergy between SOC and magnetization intrinsically generates a phase-shifted CPR~\cite{an.Joseph.SFS}.

The Josephson diode effect on the other hand is characterised by unequal critical currents in the forward and backward directions. The necessary condition for this effect is breaking of inversion and time-reversal symmetries. 

In the present work,  we demonstrate that BMS-based junctions offer a unique operational advantage: a $0$--$\pi$ transition can be induced in these systems merely by modulating the chemical potential of the BMS via an applied gate voltage. Furthermore,  we show that SOC within a BMS can drive a finite anomalous Josephson current entirely in the absence of an external magnetic field. Surprisingly, despite the breaking of both inversion and time-reversal symmetries, the junction does not exhibit a Josephson diode effect. We elucidate this physical behavior by analyzing the phase accumulated by electrons and holes during their back-and-forth traversal across the junction region.

The paper is structured as follows. Section~II details the tight-binding lattice model and Bogoliubov--de Gennes Hamiltonian governing the S--BMS--S junction. Section~III reports our results: Sec.~III\,A analyzes the critical current's dependence on the BMS band gap and chemical potential; Sec.~III\,B demonstrates the gate-tunable $0\text{--}\pi$ phase transition; and Sec.~III\,C evaluates the anomalous Josephson effect alongside the absence of superconducting nonreciprocity under spin--orbit coupling. Finally, Section~IV provides concluding remarks and discusses prospects for field-free superconducting spintronics.
 \begin{figure*}
    \centering
     \includegraphics[width=0.75\textwidth]{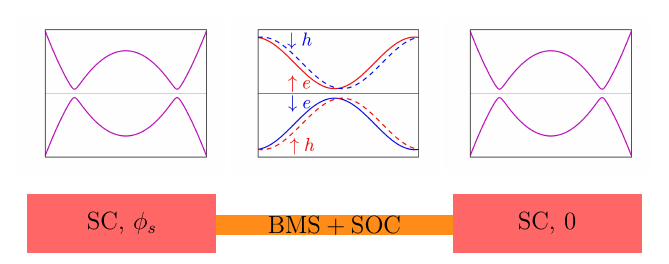}
     \caption{The schematic diagram of the set up and their respective dispersions above .}
    \label{fig:schematic}
\end{figure*}

\section{Details of calculation}
We study a junction of spin-orbit-coupled BMS sandwiched between two singlet superconductors with phase difference $\phi_s$ using lattice model [see the schematic in Fig.~\ref{fig:schematic}]. The Hamiltonian for the set up is given by $H = H_L + H_{LB} + H_{BMS} + H_{BR} + H_R$, where
\begin{equation}
\begin{aligned}
H_L
&=
-t \sum_{n=1}^{L_S-1}
\left[
\Psi_{n+1}^{\dagger}\tau_z\Psi_n
+\mathrm{h.c.}
\right]
-\mu_s \sum_{n=1}^{L_S}
\Psi_n^{\dagger}\tau_z\Psi_n
\\
&\quad
-\Delta_s \sum_{n=1}^{L_S}
\Psi_n^{\dagger}
\left(
\cos\phi_s\ \sigma_y\tau_y+\sin\phi_s\,\sigma_y\tau_x
\right)
\Psi_n ,
\\[1ex]
H_{LB}
&=
-t_j (\Psi_{L_S+1}^{\dagger}\tau_z\Psi_{L_S} + \mathrm{h.c}),
\\[1ex]
H_{BMS}
&=
-t_b \sum_{n=L_S+1}^{L_{SB}-1}
\left[
\Psi_{n+1}^{\dagger}\sigma_{z}\tau_z\Psi_n
+\mathrm{h.c.}
\right]  \\
&\quad
-\mu_b\sum_{n= L_S + 1}^{L_{SB}} \Psi_{n}^\dagger \tau_z \Psi_{n} + b \sum_{n= L_S + 1}^{L_{SB}} \Psi_{n}^\dagger  \sigma_{z} \tau_z \Psi_{n} 
\\
&\quad
-\frac{i\alpha}{2}
\sum_{n=L_s+1}^{L_{SB}-1}
\left[
\Psi_{n+1}^{\dagger}\sigma_z\Psi_n
-\mathrm{h.c.}
\right].
\\[1ex]
H_{BR}
&=
-t_j (\Psi_{L_{SB}+1}^{\dagger}\tau_z\Psi_{L_{SB}} + \mathrm{h.c}),
\\[1ex]
H_R
&=
-t \sum_{n=L_{SB}+1}^{L_{SBS}-1}
\left[
\Psi_{n+1}^{\dagger}\tau_z\Psi_n
+\mathrm{h.c.}
\right] \\
&\quad
- \sum_{n=L_{SB}+1}^{L_{SBS}}
\Big[\mu_s\Psi_n^{\dagger}\tau_z\Psi_n
+\Delta_s 
\Psi_n^{\dagger}\sigma_y\tau_y\Psi_n \Big],
\end{aligned}
\end{equation}
and  \(L_{SB}=L_S+L_B\) and \(L_{SBS}=2L_S+L_B\).
Our modeled geometry consists of a central BMS region comprising $L_B$ sites, positioned between left and right superconducting leads that each contain $L_S$ sites. At any given site $n$, the BdG spinor is defined as $\Psi_n = [c_{\uparrow,n}, \, c_{\downarrow,n}, \, c_{\uparrow,n}^{\dagger}, \, c_{\downarrow,n}^{\dagger}]^{T}$, where the operator $c_{\sigma,n}$ annihilates an electron possessing spin $\sigma$. We use $\tau_x$, $\tau_y$, and $\tau_z$ to denote the Pauli matrices operating within the particle-hole subspace. The parameters $\mu_s$ and $\mu_b$ represent the chemical potentials for the superconducting and BMS sections, respectively, while $t$ and $t_b$ correspond to their respective nearest-neighbor hopping integrals. The interface coupling between the BMS and the adjacent superconductors is dictated by the hopping amplitude $t_j$. 

To model the semiconducting band gap, we introduce the parameter $b$, which energetically separates the spin-up conduction band and the spin-down valence band. By creating an energy splitting between the two spin channels, $b$ functions as an intrinsic exchange field. Its role is remarkably similar to that of a Zeeman field, effectively lifting the spin degeneracy and determining the spin-resolved electronic structure of the material. Additionally, $\alpha$ defines the magnitude of the spin-orbit coupling inside the BMS which is taken to be along the direction of spin polarisation of BMS to get band asymmetry. The momentum-space Hamiltonian governing the bare BMS electron dispersion is formulated as
\[ H_{k} = -2t_b\cos(ka) \sigma_z + b\sigma_z + \alpha\sin(ka)\sigma_z - \mu_b\sigma_0 \]
To guarantee that a finite band gap is preserved between the two spin channels, the condition $b > \sqrt{4t_b^2+\alpha^2}$ is strictly enforced.

In standard junctions connecting two superconducting electrodes, the Josephson current is driven by subgap bound states. Because the superconducting segments in our configuration are of finite length, we determine the total supercurrent by summing the individual contributions from all occupied energy states. Due to charge conservation within the BMS region, the current operator at the interface bond between the superconductor and the BMS takes the form:
\begin{equation}
\hat{J} = -\frac{i e t_j}{\hbar} \left( \Psi^{\dagger}_{L_S+1}\Psi_{L_S} - \mathrm{h.c.} \right).
\end{equation}

To evaluate the current flow, the system's Hamiltonian is numerically diagonalized for a specified phase bias $\phi_s$, which yields the eigenstates $|u_j\rangle$ alongside their corresponding eigenenergies $E_j$. The net Josephson current is subsequently calculated using the expectation value:
\begin{equation}
J(\phi_s) = \sum_{E_j<0} \langle u_j | \hat{J} | u_j \rangle.
\end{equation}
Ultimately, the critical current ($I_c$) is extracted by finding the maximum possible supercurrent, expressed as $I_c = \max[J(\phi_s)]$.

\section{Results and Analysis}
\subsection{Josephson effect in BMS}
\begin{figure}[ht]
    \centering
    \includegraphics[width=0.9\columnwidth]{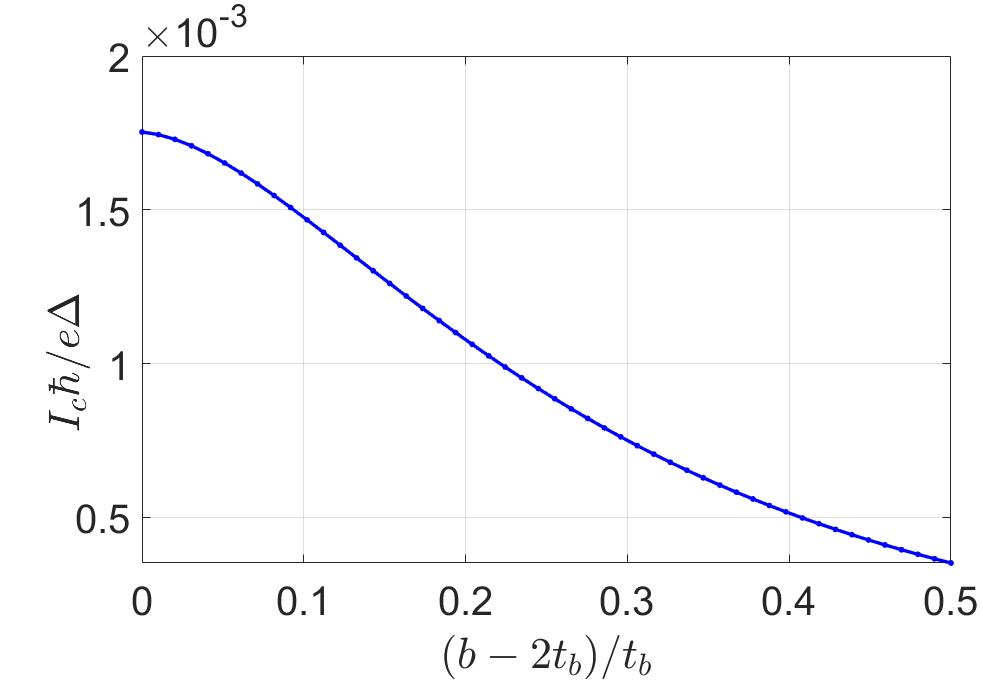}
    \caption{Critical current vs the band gap of the  BMS for the parameters $\Delta = 0.1t , \mu_s = -t ,t_j =t , t_b = t ,\mu_b = 0 , \alpha =0 ,L_S = 15 , L_B  =5.$} 
    \label{fig:Icvsb}
\end{figure}

In this section , we investigate the Josephson effect in BMS-based junctions and study the dependence of the Josephson current on key parameters, including the semiconductor band gap and the chemical potential of the BMS, setting the spin-orbit coupling strength $\al=0$. In Fig.~\ref{fig:Icvsb}, we plot the critical Josephson current as a function of the exchange field, which controls the band gap of the semiconductor, while keeping the chemical potential of the BMS set to zero. We observe that the critical Josephson current decreases with increasing band gap. This suppression can be attributed to the reduction in the low energy states  near the Fermi energy as the conduction and valence bands move farther away from the Fermi level, resulting in weaker electronic transport across the junction.  

\begin{figure}[htb]
    \centering

    \includegraphics[width=0.420\textwidth]{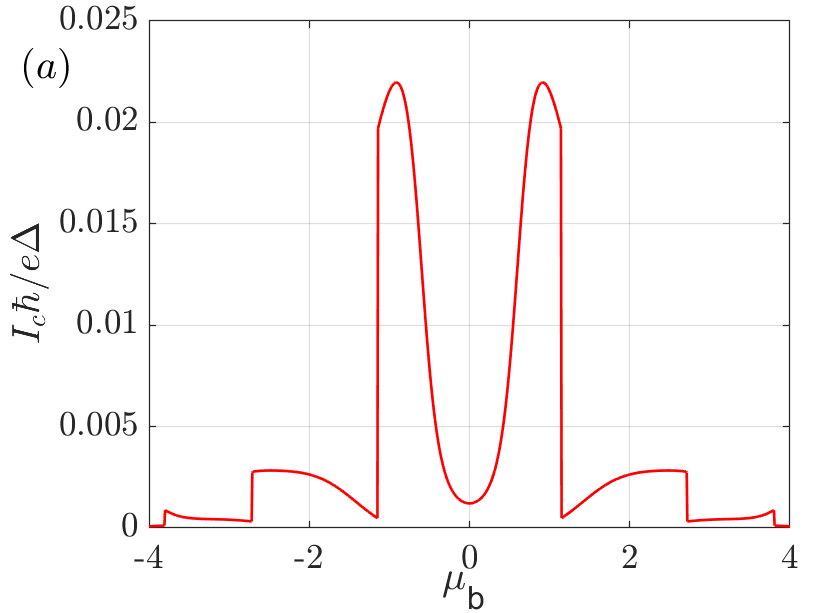}
    \includegraphics[width=0.235\textwidth]{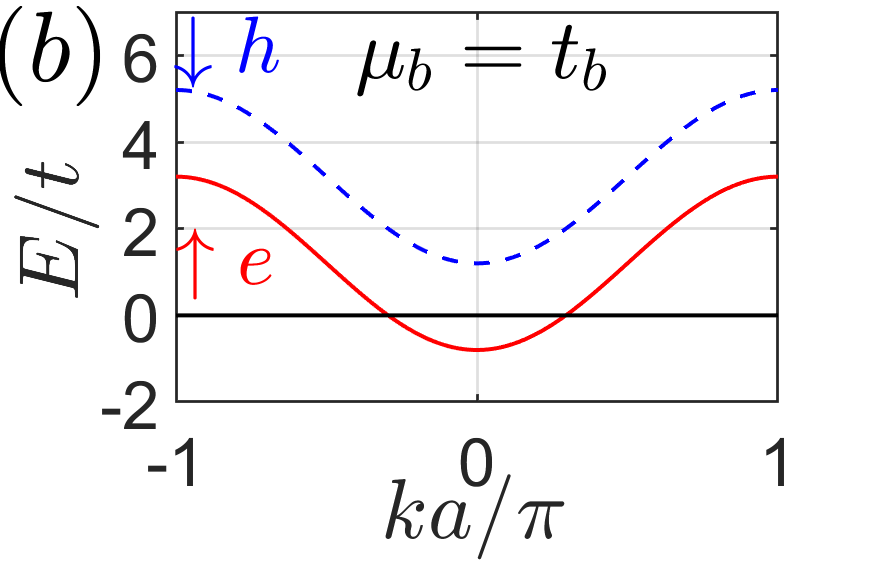}
    \includegraphics[width=0.235\textwidth]{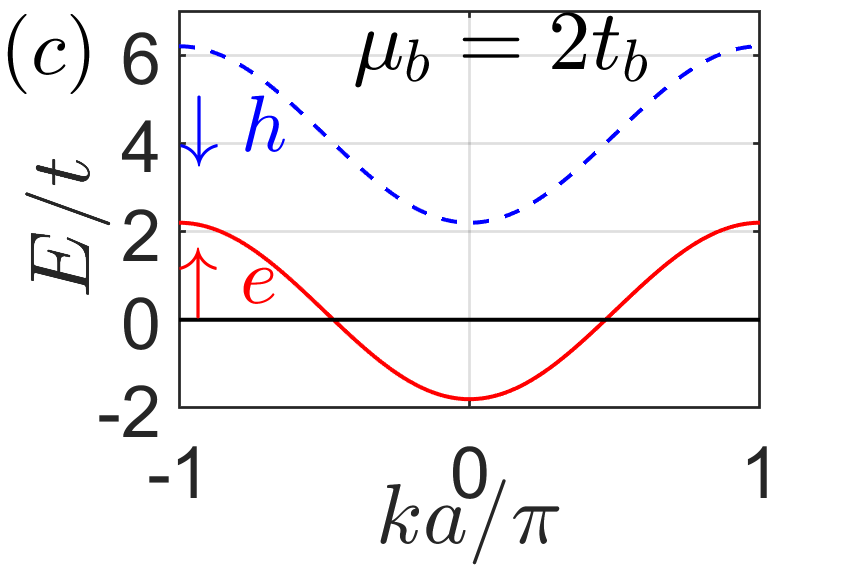}
    \caption{
    (a) Critical current versus the chemical potential of the BMS for the parameters
    $\Delta_s = 0.05t$, $t_j=t$,$t_b = t$ $\mu_s=0$, $b=2.2t_b$, $\alpha =0$, $L_S=15$, and $L_B=5$.
    Below are the dispersions of the up-spin electron and down-spin hole at
    (b) $\mu_b=t_b$ and (c) $\mu_b=2t_b$.
    }
    \label{fig:Icvsmub}
\end{figure}

We further investigated  the dependence of the critical current on the chemical potential of the BMS region [see Fig.~\ref{fig:Icvsmub}]. As the chemical potential is varied, the critical current exhibits pronounced oscillations, which can be attributed to Fabry--P\'erot interference resulting from multiple reflections of charge carriers~\cite{Sahu2023} within the BMS region. A change in the chemical potential modifies the wavevector of the propagating charge carriers and, consequently, the phase accumulated during their traversal across the junction. This leads to alternating constructive and destructive interference between the multiply reflected quasiparticle waves, producing periodic modulations in the transmission probability and, consequently, in the critical current. In addition to these interference-induced oscillations, we observe a systematic reduction in the magnitude of the critical current as the chemical potential is shifted away from the center of the BMS band gap. This suppression arises because the spin-resolved electron and hole bands move progressively farther from the Fermi energy, reducing the availability of low-energy states participating in transport. In particular, as the chemical potential is tuned away from mid-gap, the spin-up electron band and spin-down hole band become increasingly misaligned with the Fermi level [see Fig.~\ref{fig:Icvsmub}(b,c)]. Further, the wavenumber for down-spin holes is complex with larger imaginary part as $\mu_b$ is increased making it an evanescent mode.  This results in reduced quasiparticle transmission and hence, a decrease in the Josephson critical current. The minimum at $\mu_b = 0$ can be attributed to the limited availability of energy states within the superconducting gap at this particular BMS length. 

\subsection{0-\(\pi\) transitions in BMS Josephson junctions}
\begin{figure}[htb]
    \centering
    \includegraphics[width=0.45\textwidth]{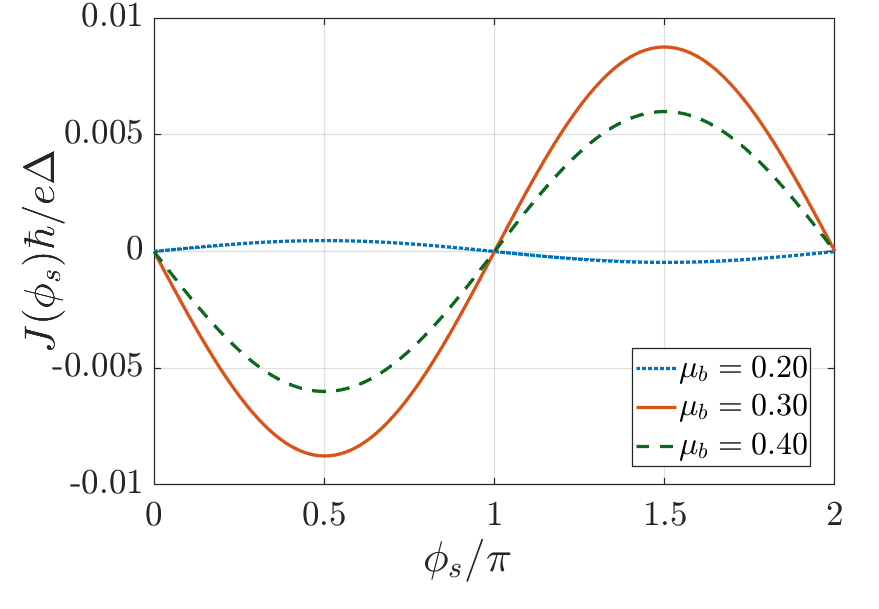}
    \caption{CPR relation for  Josephson junction of BMS for the parameters $\Delta_s =0.05t , \mu_s = -t , \ t_j = t , t_b = t,b = 2.2t_b , \alpha =0 , L_S = 10, L_B = 5$ for different values of $\mu_b/t_b $.}
    \label{fig:0topi}
\end{figure}

In this section, we investigate the $0$--$\pi$ transition  by tuning the chemical potential of the BMS region. We observe a pronounced $0$--$\pi$ transition as the chemical potential is shifted from the band-gap regime into one of the BMS bands [see Fig.~\ref{fig:0topi}]. When the chemical potential lies within the band gap, the absence of propagating states significantly suppresses the penetration of Cooper-pair correlations into the BMS region. Upon increasing the chemical potential toward the band edge and subsequently entering the conducting band, propagating states become available, leading to an enhanced penetration of superconducting correlations into the BMS region. The exchange field in the BMS region introduces a spin-dependent phase accumulation for the electron and hole components of the Cooper pairs. In particular, the spin-up electron and spin-down hole experience different phase shifts during their propagation through the BMS region. The resulting phase difference modifies the superconducting pair amplitude and produces an oscillatory behavior of the induced superconducting order parameter in the BMS region. As the chemical potential is varied, the accumulated phase changes continuously, eventually driving a reversal in the sign of the superconducting order parameter. This sign reversal corresponds to a change in the phase of the induced pair amplitude by $\pi$, thereby signaling a transition from the $0$ state to the $\pi$ state. Notably, the transition becomes possible when the chemical potential is shifted from the insulating gap into the conducting band, where propagating electronic states are available. 

\subsection{Anomalous Josephson effect in BMS }
\begin{figure}[thb]
    \centering
    \includegraphics[width=0.95\columnwidth]{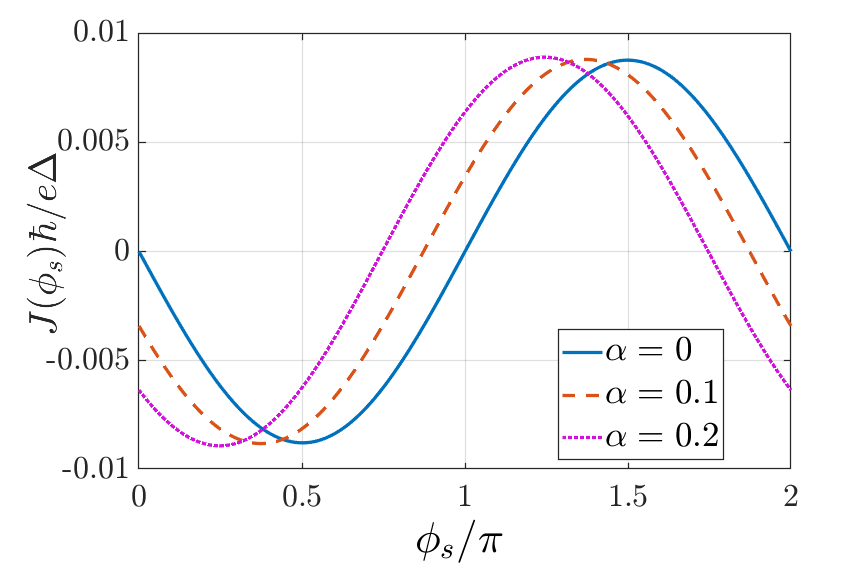}
    \caption{CPR relation for  Josephson junction of BMS with SOC for the parameters $\Delta_s =0.05t , \mu_s = -t , \ t_j =t,t_b = t, b = 2.2t_b ,\mu_b = 0.3t_b, L_S = 10, L_B = 5 . $ for different values of $\alpha/t$.}
    \label{fig:aje}
\end{figure}

In this section, we investigate the emergence of an anomalous Josephson current in a one-dimensional system exhibiting spin–orbit coupling within the BMS  region.
 we consider the configuration in which the magnetization of the BMS is aligned with the direction of the spin-orbit field since this results in breaking of inversion at the level of dispersion relation in BMS. In this configuration, spin remains a good quantum number, and the corresponding dispersions of the spin-up electron band and spin-down hole band are given by

\begin{equation}
\begin{aligned}
E_{\uparrow,e}(k) &= -2t_b\cos(ka)+ \alpha  \sin (ka) + b - \mu_b, \\
E_{\downarrow,h}(k) &= -2t_b\cos(ka) - \alpha  \sin (ka) + b + \mu_b .
\end{aligned}
\end{equation}
For up spin electron and down spin hole, 
\begin{equation}
\begin{aligned}
  k_{\uparrow , e}a = \pm \cos^{-1}\Bigg(\frac{b-E-\mu_b}{\sqrt{\alpha^2+ 4t_b^2}}\Bigg) - \delta .  \\
   k_{\downarrow , h}a = \pm \cos^{-1}\Bigg(\frac{b -E + \mu_b}{\sqrt{\alpha^2+ 4t_b^2}}\Bigg) + \delta 
  \end{aligned}
\end{equation}
 where $\delta = \cos^{-1}\big(\ 2t_b/{\sqrt{\alpha^2 + 4t_b^2}}\big)$,  the +(-) correspond to right(left) moving particles.
 The phase acquired by the right-moving spin-up electrons and left-moving spin-down holes, which carry current in the forward direction, is given by $\phi_F = (k_{\uparrow ef} - k_{\downarrow hb})L_Ba + \phi_s$. Conversely, the phase acquired by the left-moving spin-up electrons and right-moving spin-down holes, which carry current in the backward direction, is $\phi_B = (k_{\downarrow hf} - k_{\uparrow eb})L_Ba - \phi_s$. Here, $k_{\sigma ef(b)}$ and $k_{\sigma hf(b)}$ denote the wavevectors of the right(left)-moving electrons and holes, respectively, with spin $\sigma \in \{\uparrow, \downarrow\}$. When $\phi_s = 0$, we find that $\phi_F - \phi_B = 4\delta$, indicating unequal phase accumulation in the two Andreev channels. This phase asymmetry generates a finite anomalous phase shift in the CPR, leading to an anomalous Josephson effect~\cite{soori2026}. Furthermore, under the transformation $\phi_s \rightarrow 4\delta - \phi_s$, the phases $\phi_F$ and $\phi_B$ are interchanged. This symmetry ensures that for every phase acculmulated by quasiparticles that carry current in the forward direction, there exists an equivalent  phase accumulatrion by quasiparticles that carry current in the backward direction, rendering the forward and backward currents reciprocal and thereby eliminating the diode effect. The CPRs for the Josephson junction with spin-orbit coupling, which exhibits this anomalous Josephson effect, are illustrated in Fig.~\ref{fig:aje}.

\section{Summary and Conclusion}
In summary, we have shown that Josephson junctions based on  BMS provide a route to fully electrical control of supercurrents in the absence of an external magnetic field. By tuning the chemical potential of the BMS  across the band edge and into the spin-polarized regime, the junction undergoes a gate-controlled $0$--$\pi$ transition. This transition originates from the spin-dependent phase accumulation of propagating Cooper pairs and directly reflects the evolution of the spin-resolved electronic structure across the band edge. This tunability demonstrates a direct correspondence between the gate-controlled evolution of the spin-resolved band structure in the BMS region and the resulting  superconducting pairing in the junction.

We further examined the interplay between exchange-induced time-reversal-symmetry breaking and Rashba spin--orbit coupling, which breaks spatial inversion symmetry. Their combined action produces an asymmetric Andreev spectrum and a finite anomalous phase shift, $\varphi_0$, thereby displacing the equilibrium phase of the junction away from both $0$ and $\pi$ even in the absence of an external magnetic field. Notably, however, a finite $\varphi_0$ does not necessarily imply nonreciprocal Josephson transport. In the present one-dimensional BMS junction,  the energy derivatives of the phase shifts associated with the forward and backward channels remain identical. As a result, their contributions to the directional component of the supercurrent compensate exactly, preserving a reciprocal current--phase relation. Thus, although Rashba spin--orbit coupling, together with the exchange field, generates an anomalous Josephson phase, it does not lift the reciprocity of the supercurrent and is therefore insufficient, by itself, to realize a Josephson diode effect in the one-dimensional BMS Josepshson junctions considered here.

 The BMS platform consequently offers a field-free and electrically tunable means of controlling the Josephson current.  These findings establish BMS heterostructures as promising platforms for electrically controllable superconducting spintronics, phase-coherent quantum devices.
 
\begin{acknowledgments}
PN thanks Council of Scientific \& Industrial Research for financial support. AS thanks Science and Engineering Research Board (now Anusandhan National Research Foundation) Core Research grant (CRG/2022/004311) and University of Hyderabad for financial support. 
\end{acknowledgments}

\bibliography{references}

\end{document}